\documentclass[usenatbib]{mnras}
\usepackage{newtxtext,newtxmath}
\usepackage{bm,ctable,verbatim,ulem}
\newcommand{\acknowledgments}[1]{\begin{small}\section*{Acknowledgments}\end{small}{\noindent #1}\vspace{5pt}}
\newcommand{\datastatement}[1]{\begin{small}\section*{Data Availability Statement}\end{small}{\noindent #1}\vspace{5pt}}

\title[Reduced-Speed-of-Light of MHD-PIC Method]{A Reduced-Speed-of-Light
Formulation of the Magnetohydrodynamic-Particle-in-Cell Method}

\author[Ji \& Hopkins]{
\parbox[t]{\textwidth}{
Suoqing Ji$^{1,2}$ and Philip F.~Hopkins$^2$
}\vspace*{4pt} \\
$^1$ Astrophysics Division \& Key Laboratory for Research in Galaxies and
Cosmology, Shanghai Astronomical Observatory, Chinese Academy
of Sciences, \\
~80 Nandan Road, Shanghai 200030, China. suoqing@shao.ac.cn \\
$^2$ TAPIR, Mailcode 350-17, California Institute of Technology, Pasadena, CA 91125, USA. phopkins@caltech.edu \\
}

\date{}
\begin{document}
\maketitle

\begin{abstract}

A reduced-speed-of-light (RSOL) approximation is a useful technique for
magnetohydrodynamic (MHD)-particle-in-cell (PIC) simulations. With a RSOL, some
``in-code'' speed-of-light $\tilde{c}$ is set to much lower values than the true
$c$, allowing simulations to take larger timesteps (which are restricted by the
Courant condition given the large CR speeds). However, due to the absence of a
well-formulated RSOL implementation from the literature, with naive substitution
of the true $c$ with a RSOL, the CR properties in MHD-PIC simulations (e.g.\ CR
energy or momentum density, gyro radius) vary artificially with respect to each
other and with respect to the converged ($\tilde{c} \rightarrow c$) solutions
with different choices of a RSOL. Here, we derive a new formulation of the
MHD-PIC equations with a RSOL, and show that (1) it guarantees all steady-state
properties of the CR distribution function and background plasma/MHD quantities
are {\em independent} of the RSOL $\tilde{c}$ even for $\tilde{c} \ll c$, (2)
ensures that the simulation can {\em simultaneously} represent the real physical
values of CR number, mass, momentum, and energy density, (3) retains the correct
physical meaning of various terms like the electric field, and (4) ensures the
numerical timestep for CRs can always be safely increased by a factor $\sim
c/\tilde{c}$. This new RSOL formulation should enable greater self-consistency
and reduced CPU cost in simulations of CR-MHD interactions.
\end{abstract}


\section{Introduction}
\label{sec:intro}

Cosmic rays (CRs) are relativistic charged particles which can exchange energy
and momentum with the surrounding medium, and thus can potentially play a
significant role in the kinematic and thermal evolution of astrophysical fluids,
as well as being interesting in their own right as probes of high-energy
astro-particle physics. CRs are expected to be in equipartition with the
turbulent and magnetic energy density in the ISM \citep{boulares1990galactic},
and have been shown to be influential at galactic and extragalactic scales in a
number of recent numerical studies, with implications for galaxy formation
\citep{chan2019cosmic,su2019failure,hopkins2020but,buck2020effects}, galactic
outflows/winds \citep{ruszkowski2017cosmic,farber2018impact,hopkins2020cosmic},
cosmic inflows and virial shocks \citep{ji2021virial} and the phase structure of
the circumgalactic medium
\citep{salem2016role,butsky2020impact,ji2020properties,wang2020chaotic}. In
simulations of CRs on ISM or star formation or galaxy scales, as well as
classical models of Galactic CR transport which are compared to Solar system CR
experiments
\citep{strong:2001.galprop,2016ApJ...824...16J,evoli:dragon2.cr.prop}, a fluid
or Fokker-Planck type approximation for CRs is necessarily adopted, usually with
some simplified assumptions for the effective ``diffusion coefficient'' or
``streaming speed'' (or some combination thereof). The studies above, and others
comparing different assumptions for these CR ``transport coefficients''
\citep{butsky2018role,hopkins2021testing} have shown that the effects of CRs and
their interactions with the background plasma, let alone Solar system
observables, are highly dependent on these coefficients. Therefore, it is
critical to obtain a better understanding of CR transport physics and CR
``feedback'' effects (coupling to the gas via magnetic fields) at a microscopic
level.

In recent years, hybrid magnetohydrodynamic (MHD)-particle-in-cell (PIC)
simulations have been developed to investigate the interactions between CRs and
background plasma, where CRs are evolved kinetically with the PIC treatment
while the ionized gas is treated a MHD fluid
\citep{bai:2015.mhd.pic,2018ApJ...859...13M}. This is possible because, for many
astrophysical applications including all those above, the background plasma (1)
is non-relativistic (bulk speeds $u\ll c$), (2) is sufficiently well-ionized
that ideal MHD is a good approximation, and (3) has gyro radii which are vastly
smaller the CRs (e.g.\ for typical ISM protons and electrons, the gyro radii are
factors $\sim 10^{5}-10^{6}$ times smaller than the $\sim\,\mathrm{GeV}$ CRs
carrying most of the CR energy density). Compared with previous pure PIC
simulations where both CRs and ions (and sometimes electrons as well) are
simulated as particles (e.g.,
\citealt{spitkovsky2005simulations,gargate2011ion,kunz2014pegasus}), the hybrid
MHD-PIC method can simulate CR kinetic effects and feedbacks to gas by resolving
the CR gyro-radius ($\sim\mathrm{AU}$ for $\mathrm{GeV}$ CRs), without needing
to explicitly resolve the extremely tiny background plasma ion inertial length
of $\sim 10^{-6}\,\mathrm{AU}$ or even smaller background electron skin depth of
$\sim 10^{-8}\,\mathrm{AU}$. This method has been proven capable of capturing
the CR streaming instability
\citep{kulsrud.1969:streaming.instability,skilling:1971.cr.diffusion,bai:2019.cr.pic.streaming}
and accounting for the ion-neutral damping effects
\citep{kulsrud1971effectiveness,zweibel1982confinement,plotnikov2021influence,bambic2021cosmic}.
Therefore, the MHD-PIC simulations have a unique advantage in following the
evolution of CRs and their feedback/coupling to gas in many astrophysical
contexts on much greater scales and over much longer durations than pure PIC
simulations.

Even with this approximation, however, one difficulty in simulating CRs is that
CRs are relativistic with a velocity close to the speed of light $c$, far
greater than other characteristic MHD velocities (e.g., the sound speed
$c_\mathrm{s}$, the Alfv\'en speed $v_\mathrm{A}$), which severely limits the
simulation timestep (for any standard Courant-type condition) and thus makes
simulations computationally prohibitive. For instance, the speed of light
$c$ is involved in computing the Lorentz force acting on CRs as follows:
\begin{align}
    {\bm F_L} \equiv q\,[{\bm E} + ({\bm v}/c)\times{\bm B}],
    \label{eq:Lorentz}
\end{align}
where $q$ is the CR charge, ${\bm v}$ the CR velocity, ${\bm B}$ the magnetic
field (${\bm E}$ electric field), and $c$ the {\em true} speed of light possibly
with $c \gg c_\mathrm{s}, v_\mathrm{A},\text{etc.}$. For convenience, an
artificial-speed-of-light (ASOL) $\tilde{c}$ is usually used to substitute the
true speed of light $c$ in Eq.~(\ref{eq:Lorentz}), and the Lorentz factor is
accordingly rewritten as $\gamma = \tilde{c} / \sqrt{\tilde{c}^2 - \bm{v}^2}$
\citep{bai:2015.mhd.pic,2018ApJ...859...13M,van2018magnetic,Bai_2022}. With
ASOL, as long as large enough scale separations are retained, ``in-code''
simulated prpoerities can be properly rescaled to and interpreted as the
physical regimes in reality. In other contexts, such as radiation-hydrodynamics
(RHD), application of an ASOL is commonly used to allow larger timesteps for the
same reason (photons moving ``across the grid'' at $c$ require very small
numerical timesteps). But in the RHD case, since the modeling of macroscopic
systems allows little flexibility to properly rescale simulated properties,
considerable effort has gone into carefully formulating a version of the RHD
equations that ensures results are not systematically biased if the ASOL
$\tilde{c} \ll c$. In fact most current RHD formulations (and some applications
in other fields; see e.g.\ \citealt{hopkins:m1.cr.closure}) ensure that
steady-state solutions are independent of $\tilde{c}$ and that the ``true''
($\tilde{c}=c$) values of the photon energy and flux/momentum density can always
be recovered \citep[see e.g.][and references
therein]{kimm2013heavy,rosdahl2013ramses,2013ApJS..206...21S}.

However, when dealing with multi-scale problems where physical processes with
different characteristic lengths / times / energies are involved, simply
substituting $c$ with $\tilde{c}$ in MHD-PIC simulations, hereinafter refereed
to as a reduced-speed-of-light (RSOL), might cause physical inconsistencies
between simulated and ``real'' properties. For instance, since the expressions
of the CR energy density, momentum density and mass/number density contain
factors of $\tilde{c}^2$, $\tilde{c}^1$ and $\tilde{c}^0$ respectively, simply
replacing $c$ with $\tilde{c}$ for any arbitrary choice of $\tilde{c} \neq c$
makes it impossible to \emph{simultaneously} match the CR energy density,
mass/number density, \emph{and} momentum density to their correct physical
values (those with $\tilde{c}=c$). In other words, the naive substitution of $c$
in RSOL leads to changes in multiple dimensionless numbers, e.g., the CR-to-gas
energy/momentum/mass ratios, $\frac{\rho_\mathrm{cr} c^2}{ \rho_\mathrm{gas}
v_\mathrm{gas}^2} \to \frac{\rho_\mathrm{cr} \tilde{c}^2}{ \rho_\mathrm{gas}
v_\mathrm{gas}^2}$, $\frac{\rho_\mathrm{cr} c}{ \rho_\mathrm{gas}
v_\mathrm{gas}} \to \frac{\rho_\mathrm{cr} \tilde{c}}{ \rho_\mathrm{gas}
v_\mathrm{gas}}$ and $\frac{\rho_\mathrm{cr} c^0}{ \rho_\mathrm{gas}
v_\mathrm{gas}^0} \to \frac{\rho_\mathrm{cr} \tilde{c}^0}{ \rho_\mathrm{gas}
v_\mathrm{gas}^0}$, by \emph{different} factors of $(\tilde{c}/c)^2$,
$(\tilde{c}/c)^1$ and  $(\tilde{c}/c)^0$ respectively. According to the
Buckingham $\Pi$ theorem \citep{buckingham1914physically}, since characteristic
\emph{dimensionless numbers} in the simulations are changed after the native
substitution of $\tilde{c}$, simulations with and without RSOL end up with
solving totally different systems. The ``back-reaction'' forces from CRs to gas
thus can be incorrectly calculated due to the CR mass/energy/momentum density
mismatch, and the mismatch in RSOL cannot be corrected by
``post-simulation'' rescaling once the ``back-reaction'' forces have already
taken effects on-the-fly in simulations. Moreover, the CR gyro-radius also
varies with the choices of $\tilde{c}$, which leads to inconsistency in related
length scales, e.g., scales of CR scattering, in simulations with RSOL adopted.
There is no guarantee, even, with naive substitution of $c$, that the RSOL
universally ensures that the timestep can be increased by a factor $\sim
c/\tilde{c}$ (which is the goal of introducing it in the first place), nor that
steady-state solutions are independent of $\tilde{c}$.

These challenges motivate us to develop an accurate formulation of the MHD-PIC
equations with a RSOL, which resolves all of the issues above. The paper is
organized as follows. \S\ref{sec:deriv} presents the derivation of the new RSOL
formulation. We compare to previous applications of a RSOL in MHD-PIC
simulations in \S\ref{sec:compare}, discuss the numerical implementation in
\S\ref{sec:numerical}, and finally conclude in \S\ref{sec:conclude}.

\section{Derivation}
\label{sec:deriv}

Begin by considering the general Vlasov equation for a population of CRs:
\begin{align}
\frac{\partial f}{\partial t} + {\bm v} \cdot \nabla_{\bm x} f + {\bm F} \cdot \nabla_{\bm p} f = \frac{\partial f}{\partial t}{\Bigr|}_{\rm coll}
\end{align}
where $\nabla_{\bm x,\,p}$ represent the gradients in position and momentum
space, respectively, ${\bm F}$ is the force, $f$ is the distribution function,
the term on the right-hand-side represents collisional processes (which we will
neglect for now, and return to later), $t$ is time, and this is defined in the
lab (simulation) frame. We expect Lorentz forces (Eq.~\eqref{eq:Lorentz}) to
dominate in the regimes of interest, but note that our derivation is robust to
the form of ${\bm F}$.

Following the standard practice for radiation-hydrodynamics and newly-developed
methods which evolve the Fokker-Planck equation for a population of CRs
\citep{2013ApJS..206...21S,hopkins:m1.cr.closure}, we multiply this equation by
$1/c$ but then ad-hoc replace the value $1/c$ {\em only} in front of the time
derivative $\partial_{t} f$ with a different value $\tilde{c}$, to introduce the
RSOL. 
\begin{align}
\label{eqn:vlasov} \frac{1}{\tilde{c}}\frac{\partial f}{\partial t} + \frac{{\bm v}}{c} \cdot \nabla_{\bm x} f + \frac{{\bm F}}{c} \cdot \nabla_{\bm p} f = \frac{1}{c} \frac{\partial f}{\partial t}{\Bigr|}_{\rm coll}
\end{align}
This is equivalent to taking $\partial_{t} f^{\rm rsol} =
(\tilde{c}/c)\,\partial_{t} f^{\rm true}$ -- i.e.\ the time variation of $f$ is
systematically slowed by a factor of $\tilde{c}/c$, equivalent to a rescaling of
time ``as seen by'' the CRs. This is what allows us, fundamentally, to take
larger timesteps by a factor $c/\tilde{c}$, as the time variation is slower. But
this also ensures that one still recovers {\em exactly} the correct steady-state
solutions ($\partial_{t} f \rightarrow 0$) for the distribution function $f$,
independent of the choice of $\tilde{c}$. 

Now we can turn this into the equation for a ``single'' CR group. More
accurately for a MHD-PIC method which Monte-Carlo samples the dynamics of a CR
population, we assume that $f$ is a sum of $\delta$-functions each representing
a group or ``packet'' of CRs (which are represented in-code by some
``super-particles'' or other tracer field):
\begin{align}
f({\bm x},\,{\bm p},\,s,\,...) \equiv \sum_{j} N_{j}\,W ({\bm x}-\langle{\bm x}\rangle_{j})\delta({\bm p}-\langle{\bm p}\rangle_{j}) f_s(s_j),\,...
\end{align}
where $s$ denotes the species (with individual CR mass $m_{s}$), etc., $j$ and
$N_j$ is the label and total number of individual CR particles of each species
$s_j$, $W$, $\delta$ and $f_s$ are the spatial kernel weighting functions,
$\delta$-function and species function respectively, and $\langle\rangle$
represents an average of individual CR particles over a whole CR ``packet''.
Inserting this into Eq.~\eqref{eqn:vlasov} and integrating over $d^{3}{\bm
x}\,d^{3}{\bm p}\,ds...$ times ${\bm x}$, ${\bm p}$, $s$, ..., we obtain the
evolution equation for each ``packet.'' 

Ignoring collisions, we have $d_{t}{N}_{j} =0$, $d_{t}{s}_{j}=0$, $d_{t} \langle
{m}_{s}\rangle_{j}=0$, where $d_{t}{\psi}_{j} = d\psi_{j}/dt = \dot{\psi}_{j}$
represents the Lagrangian time derivative with each packet of some quantity
$\psi$, i.e.\ CR number, and individual charge, species, mass are conserved
along the trajectories of individual CRs (since we have neglected spallation and
other catastrophic processes).

For ${\bm x}$, we obtain: $\tilde{c}^{-1}\,\partial_{t}{\langle {\bm
x}\rangle_{j}}= c^{-1}\,\langle {\bm v}\rangle_{j} $, or
$\partial_{t}{\langle
{\bm x}\rangle_{j}} = (\tilde{c}/c)\,\langle {\bm v}\rangle_{j}$.  This makes
it clear how certain terms should be interpreted. Here and in all the equations
above, ${\bm v}$ represents the
{\em true} CR speed, which is an intrinsic property of the individual CR
particles. However, $\partial_{t}{ \langle {\bm x}\rangle_{j} }$ is the {\em
effective} CR advection speed of CR packets across the grid (which we can
defined as ${\bm v}_{j}^{\rm eff}$), which is reduced by our introduction of
$\tilde{c} < c$. If $|\langle {\bm v}\rangle_{j}|=c$, then trivially
$|\partial_{t}{ \langle {\bm x}\rangle_{j} } | = \tilde{c}$, so it functions
like a ``reduced speed of light'' (RSOL) as desired. The momentum equation is
(using various properties of $\delta$-functions and
cross-products to simplify the algebra):
\begin{align}
\label{eqn:momentum} \frac{1}{\tilde{c}} d_{t}{\langle {\bm p} \rangle}_{j} &= \frac{1}{c}\,\langle {\bm F} \rangle_{j} = \frac{\langle q\rangle_{j}}{c}\,\left[ {\bm E} + \frac{\langle {\bm v}\rangle_{j}}{c} \times {\bm B} \right] 
\end{align}
where ${\bm E} = -({\bm u}/c)\times {\bm B}$ to leading order here (with ${\bm
u}$ the fluid velocity, and $\rho$ the fluid density), and in the last equality
we have inserted the assumption that ${\bm F}$ comes primarily from Lorentz
forces, but our expression $d_{t} \langle {\bm p} \rangle_{j} =
(\tilde{c}/c)\,\langle {\bm F} \rangle_{j}$ is general and
allows for the introduction of any other forces (e.g.\ gravity). 

We can make this more clear by defining the following for notation purposes.
Take quantities such as the Lorentz factors $\beta\equiv |\boldsymbol{\beta}|$
and $\gamma$, and likewise the energy $E$ and momentum ${\bf p}$ of individual
CRs to have their {\em true} values, so they are intrinsic properties of each CR
particle. Then we have:
\begin{align}
{\bm v}_{j}^{\rm eff} &\equiv \partial_{t}{\langle {\bm x} \rangle}_{j} = \frac{\tilde{c}}{c}\,\langle{\bm v}\rangle_{j} = \tilde{c}\,\langle\boldsymbol{\beta}\rangle_{j} \\ 
\langle\boldsymbol{\beta}\rangle_{j} &\equiv \frac{\langle{\bm v}\rangle_{j}}{c} = \frac{{\bm v}_{j}^{\rm eff}}{\tilde{c}} \\ 
\langle\gamma\rangle_{j} &\equiv \frac{1}{\sqrt{1 - \langle\boldsymbol{\beta}\rangle_{j} \cdot \langle\boldsymbol{\beta}\rangle_{j}}} \\ 
\langle E \rangle_{j} &\equiv \langle\gamma\rangle_{j}\,\langle m_{s} \rangle_{j}\,c^{2} \\ 
\langle {\bm p}\rangle_{j} &\equiv \langle\gamma\rangle_{j}\,\langle\boldsymbol{\beta}\rangle_{j}\,\langle m_{s}\rangle_{j}\,c = \langle \gamma\rangle_{j}\,\langle m_{s}\rangle_{j}\,\langle{\bm v}\rangle_{j} = \frac{c}{\tilde{c}}\,\langle\gamma\rangle_{j}\,\langle m_{s}\rangle_{j}\,{\bm v}_{j}^{\rm eff}
\end{align}
By definition, then the {\em total} CR number, mass, momentum, and energy represented by a given ``packet'' are: $N_{j}$, $M_{j} \equiv N_{j}\,\langle m_{s}\rangle_{j}$, $\mathcal{E}_{j} \equiv N_{j}\,\langle E\rangle_{j}$, ${\bm P}_{j} \equiv N_{j}\,\langle {\bm p} \rangle_{j}$. 
We can then re-write the momentum equation for the CR ``packet'' as:
\begin{align}
\label{eqn:momentum.alt} \frac{d}{dt}\left( \langle \gamma\rangle_{j}\,\langle\boldsymbol{\beta}_{j}  \rangle \right) &\equiv \left(\frac{\tilde{c}}{c}\right)\,\left(\frac{\langle q \rangle_{j}}{\langle m_{s} \rangle_{j}\,c} \right)\,\left[ {\bm E} + \boldsymbol{\beta}_{j} \times {\bm B} \right] = \frac{1}{\tilde{c}}\,\frac{d}{dt}( \langle \gamma \rangle_{j}\,{\bm v}_{j}^{\rm eff})
\end{align}
So this is exactly the ``true'' acceleration, reduced by $\tilde{c}/c$. We now
have a momentum equation which can evolve ${\bm v}_{j}^{\rm eff}$, the effective
advection speed of a CR ``packet'' across the grid, entirely written in terms of
background MHD quantities (which retain exactly  their usual meaning) and
well-defined properties of the ``packet.''

We can derive the ``back-reaction'' force on the gas by going back to the
original Vlasov equation, calculate the force or change in momentum for
$\tilde{c}=c$, integrate over the distribution function, and apply the
equal-and-opposite change to gas, as in \citet{hopkins:m1.cr.closure}. Or,
because we are assuming tight coupling between non-relativistic ions and
magnetic fields, we could also calculate a current effect and use Ampere's law,
to get to the force on the gas, as in
\citet{zweibel:cr.feedback.review,thomas.pfrommer.18:alfven.reg.cr.transport},
and obtain the identical answer. The important thing is that just like the
radiation case, the``normal'' value of $c$ is  what appears here, which ensures
that the force on the gas/magnetic fields has its correct behavior (identical to
$\tilde{c}=c$) when the CR distribution function is in steady state. This gives:
\begin{align}
\nonumber \frac{\partial{(\rho\,{\bm u})}}{\partial t} &= -\int d^{3}{\bm p}\,f\,{\bm F} = - \sum_{j}\,n_{j}\,{\bm F}_{j} = - \sum_{j} n_{j} \langle q \rangle_{j}\,[{\bm E}+\boldsymbol{\beta}_{j}\times{\bm B}] \\ 
\nonumber &=-\sum_{j}\frac{1}{V_{j}}\, \frac{c}{\tilde{c}}\, d_{t}{\bm P}_{j} = -\sum_{j}\,n_{j}\,\frac{c}{\tilde{c}}\,d_{t}{\langle {\bm p} \rangle}_{j} \\ 
& = -\sum_{j}\,\frac{M_{j}}{V_{j}} \,\left( \frac{c}{\tilde{c}} \right)^{2}\,\frac{d}{dt}\left( \langle \gamma\rangle_{j}\,{\bm v}_{j}^{\rm eff} \right)
\end{align}
where $n_{j} \equiv N_{j} / V_{j}$ is the local CR number density with $V_{j}$
some effective differential volume assigned to the ``packet.''   

With these insights, we can also incorporate collisional effects. Returning to
Eq.~\eqref{eqn:vlasov}, ignoring the terms in ${\bm F}$, we simply have $d_{t}
\psi_{j} = (\tilde{c}/c)\,d_{t} \psi_{j} |_{\rm coll}$ for some $\psi$ and
relevant collisional effect (e.g.\ catastrophic losses). So for example if
catastrophic losses remove CRs at a ``true'' rate $\dot{n}_{j} = -\langle \sigma
v \rangle_{j}\,n_{n}\,n_{j}$ (where $n_{n}$ is the local nucleon number
density), the total number/mass/momentum/energy of a ``packet'' is reduced
accordingly as $d_{t} \psi_{j} \rightarrow -(\tilde{c}/c)\,\langle \sigma v
\rangle_{j}\,n_{n} \,\psi_{j}$ for
$\psi_{j}=(N_{j},\,M_{j},\,\mathcal{E}_{j},\,{\bm P}_{j})$. 

Upon examination, a few points are clear. First, as intended, the manner in
which $\tilde{c}$ enters is equivalent to a rescaling of time as seen by the
CRs, so it guarantees that we can always increase the timestep by $c/\tilde{c}$.
This means that the gyro-frequency is reduced in the lab frame (necessarily) by
the same factor (as $\Omega_{g}^{\rm eff} = \Omega_{g}\,(\tilde{c}/c)$), so the
actual {\em effective} time it requires to complete a gyro orbit is longer.
However, this means that for a given true ${\bm v}_{j}$ or equivalently CR
energy or $\gamma_{j}$, the gyro {\em radius} on  the grid is entirely
independent of $\tilde{c}$. We can explicitly confirm either if we think about
the RSOL as a rescaling in time, or simply examine the ``effective'' transport
speeds across the grid, that the Courant criteria for CR ``advection'' (gyro
orbits) become e.g.\ $\Delta t < C/\Omega_{g}^{\rm eff} = (c/\tilde{c})\,\Delta
t(\tilde{c}=c)$; likewise for advection ``through a cell'' (which restricts both
CR and MHD timesteps), $\Delta t < C\,\Delta x / {\bm v}^{\rm eff} =
(c/\tilde{c})\,\Delta t(\tilde{c}=c)$. Just like with radiation-hydrodynamics,
the ``conserved momentum'' here upon MHD-CR exchange is $\rho\,{\bm u} +
(c/\tilde{c})\,{\bm P}$ (where ${\bm P} \equiv \sum_{j} P_{j}$), not $\rho\,{\bm
u} + {\bm P}$ (which is only true if $\tilde{c}=c$). From the form of the Vlasov
equation and force on gas written in terms of the distribution function $f$, it
is also immediately obvious that so long as the {\em distribution function is in
steady-state}, i.e.\ $\partial_{t} f \rightarrow 0$, then we are guaranteed to
recover the correct steady-state solutions for the distribution function, i.e.\
the correct steady state distribution of CR velocities, phase and pitch angles,
momenta/Lorentz factors, etc, as well as the correct force on the gas. We are
also guaranteed that processes such as scattering occur on the correct length
scales: e.g.\ scattering rates (or e.g.\ collisional/catastrophic loss rates)
will be slowed by a factor $\tilde{c}/c$, but so will transport speeds, so by
the time a CR has traveled a given {\em distance} $\ell$ down a magnetic field
line, it will have an identical integrated scattering/catastrophic loss
probability, independent of $\tilde{c}$.

\section{Comparison To Naive RSOL Substitution}
\label{sec:compare}

While ``pure PIC'' methods are quite mature, MHD-PIC methods for applications
such as CR dynamics have only been widely-developed relatively recently. As
previously discussed, the ASOL method can be interpreted as changing the unit
system while retaining the same dimensionless numbers of a problem. While when
dealing with multi-scale problems where prohibitively small timestep might
become an issue, simply treating the RSOL $\tilde{c}$ as if the ``true'' SOL $c$
were reduced for the CRs specifically, does enable much larger timesteps, but
leads to several conceptual difficulties and inconsistencies in an effort to
match the physical properties in real systems, including: (1) it is impossible
to simultaneously reproduce the true CR energy and/or momentum and/or number
densities, regardless of whatever limit the system is in, since for example in
their formulation $e_{j} \equiv d\mathcal{E}_{j}/d^{3}{\bm x} =
n_{j}\,\gamma_{j}\,\langle m_{s}\rangle_{j}\,\tilde{c}^{2}$, so either $n_{j}$
or $e_{j}$ must be incorrect (and likewise for momentum density); (2) the
transition between relativistic and ultra-relativistic behavior occurs at the
incorrect CR energy/momentum; (3) this leads to a momentum equation (in our
notation) $d_{t}(\langle \gamma \rangle_{j}\,{\bm v}_{j}^{\rm eff}) \rightarrow
(\langle q \rangle_{j}/\langle m_{s}\rangle_{j}\,c)\,(c\,{\bm E} + {\bm
v}_{j}^{\rm eff} \times {\bm B})$, which is different from
Eq.~\eqref{eqn:momentum.alt} by $(c/\tilde{c})^{2}$ in the electric-field term
and $(c/\tilde{c})^{1}$ in the magnetic-field term, breaking the physical
correspondence between the two unless one redefines the electric field ${\bm E}$
to be different from that assumed in the MHD derivation; (4) likewise if one
rescales this momentum equation to keep $(q/m_{s}\,c)$ fixed, then this leads
not to a rescaling of $\Omega$ with $\tilde{c}$, but gives the same $\Omega$
(requiring the same timestep as simulations with $\tilde{c}=c$, for gyro
orbits), instead with a smaller gyro radius $\tilde{c}\,\Omega$; (5) there is no
guarantee that when the distribution function is in steady-state (even for the
simplest homogeneous gas and CR configuration), the distribution function and/or
the net back-reaction force on the gas will actually be correct, compared to
their value with $\tilde{c}=c$. As a result, in these approaches, one obtains
certain results that vary systematically with $\tilde{c}$ and must be
extrapolated (e.g.\ by running a large suite of simulations with different
$\tilde{c}$) to estimate their correct $\tilde{c} \rightarrow c$ values. In
contrast, our approach allows for formal convergence of steady-state properties
with $\tilde{c} \ll c$. 

We finally note that our proposed RSOL formulation actually {\em derives} the
``correction terms'' which are needed to add to the electric field, CR
acceleration equation, and ``back-reaction'' force equation, and generically
guarantees that the CR distribution function will give the correct steady-state
answer. However, simply replacing $c$ with RSOL $\tilde{c}$, by definition, will
{\em not} solve the correct CR distribution function in steady state. The
previous method is equivalent to a simultaneous rescaling of both the velocities
and spatial units of the original problem, so by decreasing $c$, it ends up with
solving a \emph{different} problem from what is intended to solve originally,
specifically a problem with a different range of spatial scales and a different
ratio of $v_\mathrm{A}/c$, $c_s/c$ and $u/c$. Whereas our proposed method,
equivalent to a rescaling of time (``as seen by'' the CRs) only, can obtain the
right answer for the {\em actually desired} dimensionless numbers of the problem
simultaneously (and all physical properties as well, e.g., CR mass, momentum and
energy densities), so long as the CR distribution function is in steady-state.
Therefore, whenever a RSOL is needed for enabling larger timesteps, our proposed
RSOL method should always be implemented in order to achieve the best possible
consistency within simulated systems.

\section{Numerical Implementation \&\ Timestep Constraints}
\label{sec:numerical} 

The form of the RSOL proposed here can be \emph{immediately and trivially}
implemented in any MHD-PIC code with minimal effort, as it amounts to a
straightforward renormalization of the time-derivative terms which would be
computed anyways. We have explored existing implementations in the code {\small
GIZMO} in \citet{ji2022numerical}, but note it immediately applies to any other
MHD-PIC implementation, for example both the ``full-$f$'' and ``$\delta f$''
methods in \citet{bai:2015.mhd.pic,bai:2019.cr.pic.streaming}. In these methods
the MHD (non-relativistic ion/neutral/electron) dynamics are solved in some
fluid limit with whatever solver is most useful, additional physics (e.g.\
radiation transport or dust as in \citealt{ji:cr.dust}) can be added as well,
and then the CRs are integrated on this background with a ``super-particle''
approach. In this, one simply identifies each super-particle with a ``packet''
above, and immediately obtain their evolution equations. The only difference
between our approach and one with no RSOL is the insertion of appropriate
factors of $\tilde{c}/c$ in the code. 

With this formulation, we see also that when initializing the system, one should
simply initialize the correct $N_{j}$ such that e.g.\ $N \equiv \sum{N}_{j} =
M_{\rm cr}^{\rm total} / \langle m_{s} \rangle$, where $\langle m_{s} \rangle$
is the true individual CR mass. Or equivalently $N = E_{\rm cr}^{\rm total} /
\langle E\rangle_{j} = E_{\rm cr}^{\rm total} /
\langle\gamma\rangle_{j}\,\langle m_{s} \rangle_{j}\,c^{2}$. This ensures that
the correct total CR energy and momentum densities $\mathcal{\epsilon}$ and
${\bm P}$ are also initialized -- these are all automatically consistent with
one another. Inconsistency only arises if we incorrectly identify
$\langle\gamma\rangle_{j}\,\langle m_{s}\rangle_{j}\,{\bm v}_{j}^{\rm eff}$ as
the individual CR momentum, but this is not correct because the relation between
${\bm v}_{j}^{\rm eff}$ and the momentum as it appears in conservation and
dynamics equations is changed by the introduction of the RSOL. As detailed
above, quantities like ${\bm p}$, ${\bm v}$, $\gamma$, $\boldsymbol{\beta}$,
$E$, etc.\ all retain their usual intrinsic physical meaning (that they would
have with $\tilde{c}=c$), but the actual advection velocity $\partial_{t}\langle
{\bm x} \rangle_{j}$ of a ``super-particle'' is ${\bm v}_{j}^{\rm eff}$, and the
CR momentum equation is modified with the appropriate $\tilde{c}/c$ factors. 

The result of this is that all timestep constraints {\em for the CRs} also
rescale by $c/\tilde{c}$, as all the time-derivative terms for CRs rescale with
this. Obviously, one requires $\Delta t_{j} < C\,\Omega_{j}^{-1} = C\,\langle
r_{\rm g}\rangle_{j}/\tilde{c} = (c/\tilde{c})\,C\,(\langle r_{\rm g}\rangle_{j}
/ c)$, where $\langle r_{\rm g}\rangle_{j}$ is the gyro radius, and $C$ is a
Courant factor, or more generally $\Delta t_{j} < C \,| {\bm v}_{j}^{\rm eff} |
/ | d_{t} {\bm v}_{j}^{\rm eff} |$, and if there are some catastrophic losses,
similarly $\Delta t_{j} < C\, N_{j}/|\dot{N}_{j}|  \sim  (c/\tilde{c})/(\langle
\sigma v \rangle_{j}\,n_{n})$. And for the Courant condition of CRs propagating
through the gas with some cell size $\Delta x$, we have $\Delta t_{j} <
C\,\Delta x/|\partial_{t} {\bm x}|_{j} = C\,\Delta x / |{\bm v}_{j}^{\rm eff}| =
(c/\tilde{c})\,C\,\Delta x / | \langle {\bm v}\rangle_{j}|$. For the case with
non-relativistic CRs, even though it does not gain significant benefit from the
RSOL regarding computational time saving, this RSOL implementation is still
recommended: it naturally returns to the ASOL formulation under such regimes,
and can be safely and conveniently extrapolated to relativistic regimes where a
RSOL might be needed.

Finally, just like any other RSOL implementation, convergence to correct
solutions requires choosing a sufficiently large value of $\tilde{c}$ compared
to other (e.g.\ MHD) signal speeds in the problem. Although formally
steady-state solutions are independent of $\tilde{c}$ for all $\tilde{c}$ with
this method, if one chose $\tilde{c}$ too small compared to e.g.\ some rms
turbulent velocity of the MHD fluid, $\tilde{c} \ll |\delta u|$, then the plasma
would artificially ``outpace'' the CRs and the system will not actually be able
to converge to a steady-state distribution function $f$ in the first place. This
is also consistent with the requirement of sufficient scale separation in order
to ensure, e.g., a much smaller gyro-period than the eddy turnover / Alfv\'en
crossing timescales.

\section{Concluding Remarks}
\label{sec:conclude}

In this paper, we propose an accurate, well-defined formulation of a RSOL which
can be applied to MHD-PIC simulations. The major advantages of the proposed
formulation are that: (1) The timestep can be safely increased by $c/\tilde{c}$
as intended (regardless of if the timestep restriction from gyro orbits,
advection over the grid, grid response to CRs, or collisional terms dominates).
But meanwhile (2) the key properties of CRs in numerical simulations, such as
energy density, momentum density, mass/number density and CR gyro-radius, are
\emph{independent} of the choices of $\tilde{c}$, and (3) they can
\emph{simultaneously} reproduce their true, physical values (those one would
have with $\tilde{c}=c$). And perhaps most importantly (4) solutions for CR bulk
properties, distribution function, and their interactions with/forces on the
magnetic fields and background fluid are independent of $\tilde{c}$ as well
(even if $\tilde{c} \ll c$) so long as the CR distribution function $f$
approaches a local steady-state. These advantages naturally recommend the
formulation here as a standard and straightforward implementation for MHD-PIC
simulations whenever a RSOL is required. 

\datastatement{There are no new data associated with this article.}

\acknowledgments{The authors thank the referee and editor for their constructive
suggestions which improve this work. SJ is supported by a Sherman Fairchild
Fellowship from Caltech, the Natural Science Foundation of China (grants
12133008, 12192220, and 12192223) and the science research grants from the China
Manned Space Project (No. CMS-CSST-2021-B02). Support for PFH was provided by
NSF Research Grants 1911233 \&\ 20009234, NSF CAREER grant 1455342, NASA grants
80NSSC18K0562, HST-AR-15800.001-A. Numerical calculations were run on the
Caltech compute cluster ``Wheeler'', allocations FTA-Hopkins supported by the
NSF and TACC, and NASA HEC SMD-16-7592, and the High Performance Computing
Resource in the Core Facility for Advanced Research Computing at Shanghai
Astronomical Observatory.}

\bibliography{ms_extracted}

\end{document}